# High-$T_c$ Superconductivity in some Heavy Rare-earth Iron-arsenide REFeAsO$_{1-\delta}$ (RE = Ho, Y, Dy and Tb) Compounds

**Jie Yang[+], Xiao-Li Shen[+], Wei Lu, Wei Yi, Zheng-Cai Li,**
**Zhi-An Ren[*], Guang-Can Che, Xiao-Li Dong, Li-Ling Sun, Fang Zhou, Zhong-Xian Zhao[*]**

National Laboratory for Superconductivity, Institute of Physics and Beijing National Laboratory for Condensed Matter Physics, Chinese Academy of Sciences
P. O. Box 603, Beijing 100190, P. R. China

Abstract:
  New iron-arsenide superconductors of REFeAsO$_{1-\delta}$ (RE = Ho, Y, Dy and Tb) were successfully synthesized by a high pressure synthesizing method with a special rapid quenching process, with the onset superconducting critical temperatures at 50.3 K, 46.5 K, 52.2K and 48.5 K for RE = Ho, Y, Dy and Tb respectively.

Recently new high temperature superconductivity was discovered in some iron-arsenide compounds REFeAsO$_{1-x}$F$_x$ (RE = rare earth elements) by fluorine doping, with the highest critical temperature ($T_c$) at 55 K [1-6]. With the FeAs layer was found to be responsible for the unconventional superconductivity inside, F-free while oxygen-deficient REFeAsO$_{1-\delta}$ compounds were also discovered to be superconducting at similar temperatures [7]. These quaternary superconductors have a layered tetragonal ZrCuSiAs type structure with space group P4/nmm (denoted as 1111 phase), and can be described by the alternating stacking of [Fe$_2$As$_2$]$^{2+}$ layers and [RE$_2$O$_2$]$^{2-}$ layers. Besides, some oxygen-free iron-arsenide superconductors were also discovered later, such as AFe$_2$As$_2$ (A = Ba, Sr, Ca and Eu, *etc.*) by carrier doping or under pressure, Li$_x$FeAs, FeSe *etc.*, which superconducts at lower temperatures [8-14]. Till now the electron pairing mechanism is still unclear in these superconducting materials, while all these iron-arsenide superconductors have made up the second high-$T_c$ family after cuprates.

After the discovery of F-free REFeAsO$_{1-\delta}$ superconductors with RE = La, Ce, Pr, Nd and Sm [7], we have also investigated these quaternary compounds of all other heavy rare earth elements by a high pressure synthesis method (HP), and clear superconducting signals have been observed in some multi-phase samples with the $T_c$ around 50 K when RE = Gd, Tb, Dy, Ho and Y. However, the superconducting phase can not be distinguished by X-ray diffraction (XRD) analysis except for Gd [15]. Late Scanning Tunneling Microscopy investigation indicated the nanometer size of superconducting grains, which are not sensitive to XRD. A possible reason is that the 1111 phases with these heavy RE elements are meta-stable and easy to decompose. Independently, by synthesizing under a ultra-high pressure of about 12 GPa, F-doped 1111 phase of TbFeAs(O,F) and DyFeAs(O,F) were reported to be superconducting around 46 K [16].

Here we report the high-$T_c$ superconductivity in some oxygen-deficient REFeAsO$_{1-\delta}$ (RE = Ho, Y, Dy, Tb) compounds, which were successfully synthesized by a revised HP method with a rapid quenching approach.

The starting chemicals of Ho, Y, Dy, Tb chips, and As, Fe, $Fe_2O_3$ powders are all with the purity better than 99.99%. At the first step, REAs powder was obtained by reacting RE pieces and As powders at 650 °C for 12 hours and 1180 °C for 40 hours, and then thoroughly grounded into fine powder after cooled. For the HP synthesis, the appropriate amount of REAs, Fe and $Fe_2O_3$ powders was mixed together according to a nominal chemical formula $REFeAsO_{0.8}$, then grounded thoroughly and pressed into small pellets. The pellet was sealed in boron nitride crucibles and then mounted in a six-anvil high pressure synthesis apparatus. After applied a high pressure of 5 GPa, the temperature was increased to 1000 °C by electrical heating, and sintering for 1 hour while keeping the pressure constant. Here a special rapid cooling process was adopted. After switching off the power, the sample was quickly quenched to room temperature by water-cooling in about 1 minute and then the pressure was released (HP method). The rapid quenching process was supposed to keep the meta-stable phase that formed under high temperature and high pressure. The synthesized samples are dense and hard with black color. For comparison, we have also synthesized the above mentioned REFeAsO undoped compounds in sealed evacuated quartz-tube at 1180 °C for 60 hours (AP method).

The phase purity and structural identification were characterized by powder X-ray diffraction (XRD) analysis on an MXP18A-HF type diffractometer with Cu-$K_\alpha$ radiation from 20° to 80° with a step of 0.01°. The XRD results indicate that for the 1111 phase of the AP samples, only TbFeAsO compound was obtained, with the lattice parameters a = 3.898(1) Å, c= 8.404(2) Å, while other rare earth elements cannot form the parent 1111 phase by the AP method. The XRD pattern of the AP TbFeAsO sample was plotted in Fig. 1, together with a calculated pattern for identification of the 1111 phase. For the HP samples, besides the reported 1111 phases for RE = Dy and Tb, the same phases for RE = Ho and Y were discovered here. The XRD patterns for $HoFeAsO_{1-\delta}$ and $YFeAsO_{1-\delta}$ were also plotted in Fig. 1, and the main diffraction peaks were pointed out by arrows. For all these samples with heavy rare earth metals, impurities always exist in the synthesized samples, and the main impurity phases were identified to be $RE_2O_3$, REAs and FeAs, which are all not superconducting in the measuring temperature. These

REFeAsO$_{1-\delta}$ phases were speculated to be meta-stable and decomposed at a lower temperature, and this explains why the quenching process worked to preserve the 1111 phase. The lattice parameters of all these synthesized 1111 phases were summarized in Table. 1. Clearly, with the increase of ionic size of the RE elements (Ho $\leqslant$ Y < Dy < Tb), the lattice parameters monotonously increase; and oxygen-vacancy causes the decrease of lattice parameters as we reported.

Temperature dependence of the resistivity for all samples was measured by a standard four-probe method from 5 K to 300 K. The AP samples were not superconducting while all HP samples exhibit superconducting zero resistance at low temperatures above 40 K (as shown in Fig. 2), with nearly linear resistivity vs. temperature behavior above the superconducting transition, similar with the previous discovered 1111 superconductors with light rare earth elements. The onset superconducting transition temperatures ($T_c$(onset)) for these REFeAsO$_{1-\delta}$ superconductors are 50.3 K, 46.5 K, 52.2K and 48.5 K for RE = Ho, Y, Dy and Tb respectively, as shown in Table. 1.

The DC magnetization was measured using a Quantum Design MPMS XL-1 system. For an experiment cycle the sample was cooled to 1.8 K in zero field cooling (ZFC) and data were gathered when warming in an applied field, then cooled again under an applied field (FC) and measured when warming up. The DC-magnetization data (measured under a magnetic field of 1 Oe) for all of the HP samples of HoFeAsO$_{1-\delta}$, YFeAsO$_{1-\delta}$, DyFeAsO$_{1-\delta}$ and TbFeAsO$_{1-\delta}$ are shown in Fig. 3 with both of ZFC and FC results, with the onset diamagnetic transition temperature (determined from the ZFC curves) at 49.1 K, 45.2 K, 51.0 K, and 46.8 K respectively. These are a little lower than the $T_c$(onset) determined from the resistivity measurements, which can be attributed to the poorer quality of these superconductors than those with light rare earth elements [7], because of the difficulty in forming 1111 phases. Nevertheless, the diamagnetic superconducting shielding fractions for all these superconductors are from 30% to 75%, which indicate the bulk superconducting behavior of these compounds.

In conclusion, superconductivity has been achieved in some layered quaternary REFeAsO$_{1-\delta}$ compounds (for RE = Ho, Y, Dy and Tb) by the introducing of oxygen vacancy through HP synthesis, and the 1111 phase structure was successfully stabilized by a rapid quenching process. Therefore, together with previous discovery, high-$T_c$ superconductivity has been discovered in all of these REFeAsO$_{1-\delta}$ compounds for RE = La, Ce, Pr, Nd, Sm, Gd, Tb, Dy, Ho and Y.

Acknowledgements:

We thank Mrs. Shun-Lian Jia for her kind helps in resistivity measurements. This work is supported by Natural Science Foundation of China (NSFC, No. 50571111 & 10734120) and 973 program of China (No. 2006CB601001 & 2007CB925002). We also acknowledge the support from EC under the project COMEPHS TTC.

[+] Authors with equal contributions;
[*] Corresponding Authors:
Zhi-An Ren: renzhian@aphy.iphy.ac.cn
Zhong-Xian Zhao: zhxzhao@aphy.iphy.ac.cn

Figure Captions:

Figure 1: Typical XRD patterns for synthesized undoped TbFeAsO (AP), superconducting YFeAsO$_{1-\delta}$ and HoFeAsO$_{1-\delta}$ (HP) compounds; the vertical bars indicate the calculated diffraction peaks for the TbFeAsO.

Figure 2: The temperature dependences of resistivity for the REFeAsO$_{1-\delta}$ (RE = Ho, Y, Dy and Tb) superconductors synthesized by HP method.

Figure 3: The temperature dependence of the DC-susceptibility (H = 1Oe) for the REFeAsO$_{1-\delta}$ (RE = Ho, Y, Dy and Tb) superconductors synthesized by HP method.

Table 1. The lattice parameters and superconducting critical temperature for some REFeAsO$_{1-\delta}$ compounds (RE = Ho, Y, Dy and Tb)

| REFeAsO$_{1-\delta}$ | a / Å | c / Å | Vol. / Å$^3$ | $T_c$ (onset-R) / K |
|---|---|---|---|---|
| HoFeAsO$_{1-\delta}$ | 3.846(2) | 8.295(3) | 122.7 | 50.3 |
| YFeAsO$_{1-\delta}$ | 3.842(3) | 8.303(5) | 122.6 | 46.5 |
| DyFeAsO$_{1-\delta}$ | 3.859(1) | 8.341(4) | 124.2 | 52.2 |
| TbFeAsO$_{1-\delta}$ | 3.878(2) | 8.354(3) | 125.7 | 48.5 |
| TbFeAsO | 3.898(1) | 8.404(2) | 127.7 | - |

Figure 1:

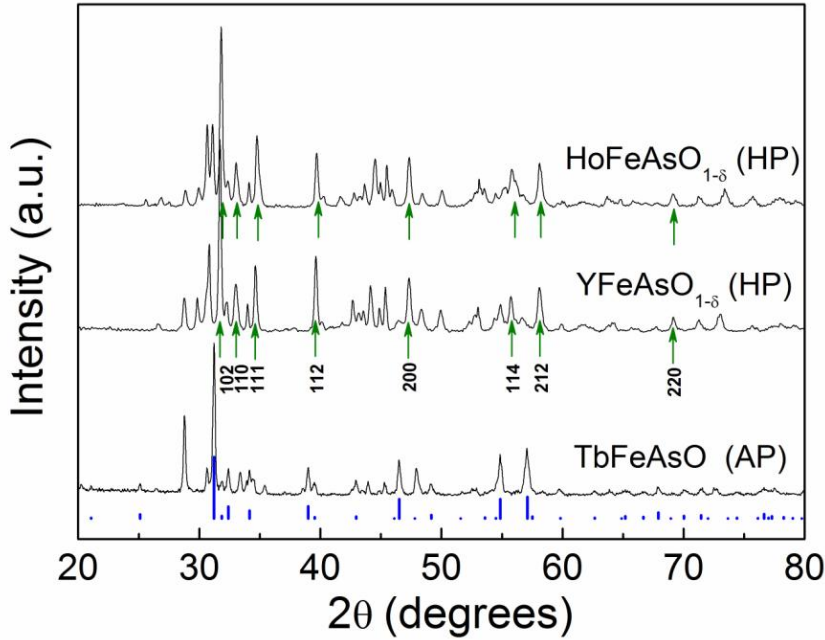

Figure 2:

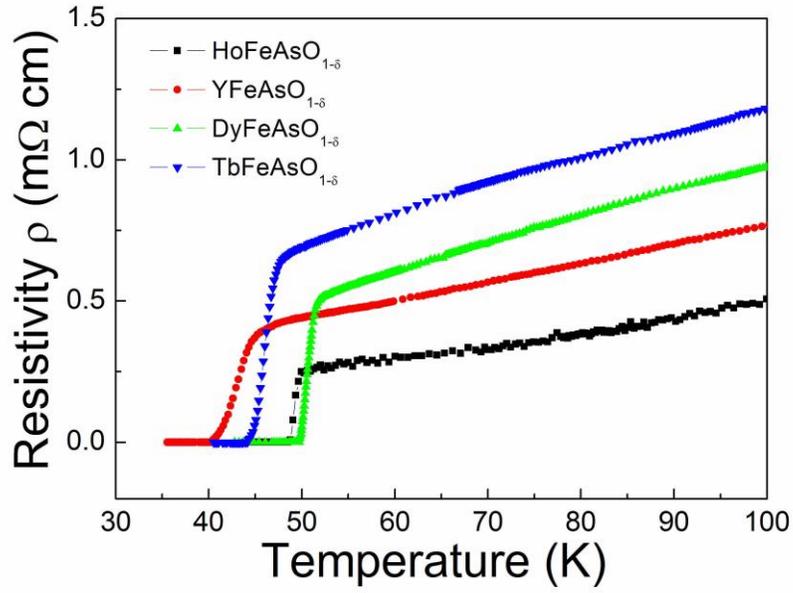

Figure 3:

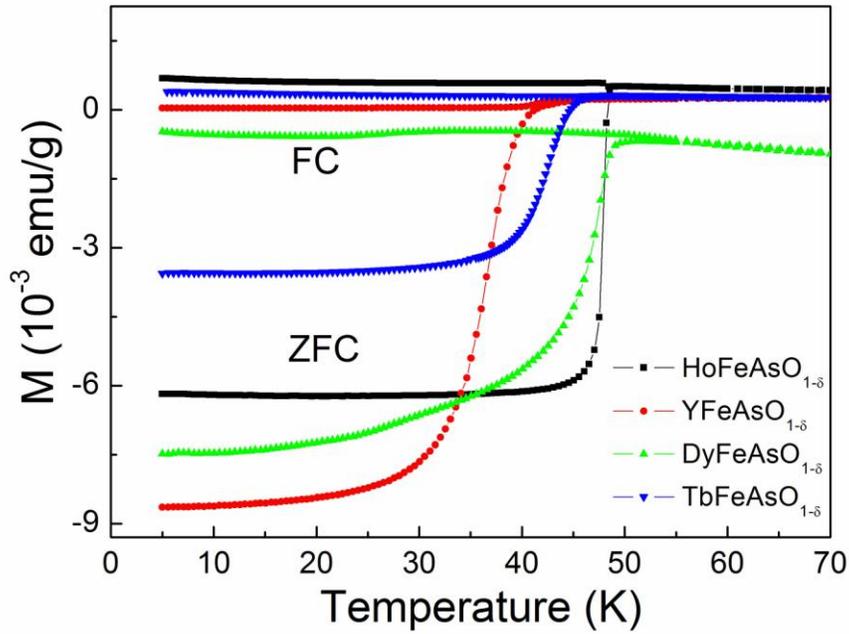